\documentclass[aps,pra,twocolumn,showpacs,superscriptaddress]{revtex4-2} 
\usepackage{graphicx}
\usepackage{amsmath}
\usepackage{amssymb}
\usepackage{epstopdf}
\usepackage{amsfonts}
\usepackage{dcolumn}
\usepackage{bigints}
\usepackage{xcolor}
\usepackage{braket}
\usepackage[normalem]{ulem}
\usepackage{comment}

\begin{document}
\title{Tunable Non-Gaussian Mechanical States in a \\ Strongly Coupled Hybrid Quantum System}
\author{Jugal Talukdar}
\email{jug.talukdar@gmail.com}
\address{Division of Physical Sciences, College of Letters and Science, University of California, Los Angeles, CA 90095 USA}
\author{Scott E. Smart}
\address{Division of Physical Sciences, College of Letters and Science, University of California, Los Angeles, CA 90095 USA}
\author{Prineha Narang}
\email{prineha@ucla.edu}
\address{Division of Physical Sciences, College of Letters and Science, University of California, Los Angeles, CA 90095 USA}
\address{Department of of Electrical and Computer Engineering, University of California, Los Angeles, CA 90095 USA}

\date{\today}

\begin{abstract}
Quantum states of motion are critical components in the second quantum revolution. We investigate the generation and control of non-Gaussian motional states in a tripartite hybrid quantum system consisting of a collection of qubits coupled to a mechanical resonator, which in turn interacts with an externally driven photonic cavity. This hybrid architecture provides a versatile platform for quantum control by integrating nonlinear interactions and multiple control parameters. Operating in the strong coupling regime, we study the transient dynamics resulting from a time-dependent external drive that has a boxcar profile. Starting from coherent states in both the mechanical and cavity subsystems, we show that this drive protocol, combined with time-independent interaction and frequency configurations, leads to the emergence of highly non-Gaussian quantum states in the intermediary mechanical degree of freedom. These states are characterized by a pronounced negative volume in the Wigner quasi-probability distribution and enhanced quantum Fisher information, indicative of their quantum utility. We systematically analyze the impact of the qubit phase, interaction strengths, and drive parameters on the degree of non-Gaussianity. Our findings underscore the tunability and richness of this hybrid platform, paving the way for advanced quantum state engineering and applications in quantum sensing, metrology, and information processing. 
\end{abstract}
\maketitle

\textit{Introduction:} Tunable and adaptable continuous variable quantum states are essential to harness the full potential of physical systems operating in the quantum regime~\cite{ref_Duan,ref_Lloyd,ref_Iosue,ref_Chabaud}. These states have not only expanded our understanding of quantum mechanics~\cite{ref_Chemmn} but also enabled quantum-enhanced phenomena with promising practical applications~\cite{ref_Lvovsky,ref_Lance,ref_Msdsen,ref_Regula,ref_Marshall,ref_Menicucci,ref_braunstein,ref_zhanggg,ref_Cardinal,ref_Furrer,ref_Pirandola,ref_Nokkala}. As we navigate the second quantum revolution, the importance of quantum states possessing non-trivial correlations and characteristics beyond the realm of Gaussian states has become increasingly evident~\cite{ref_walschaers_0,ref_walschaers,ref_ra,ref_khalili,ref_mendo,ref_genoni,ref_stel-rank,ref_genoni_1,ref_cert,ref_straka,ref_park_1}. Highly non-Gaussian quantum states, such as Gottesman-Kitaev-Preskill (GKP) states, are promising candidates for efficient quantum information encoding and quantum error correction~\cite{ref_gkp, ref_marek_2, ref_noh_1}. Successful experimental demonstrations of GKP states~\cite{ref_Lachance-Quirion, ref_Kolesnikow, ref_Dahan, ref_Konno, ref_Hastrup, ref_Bohnmann, ref_Pizzimenti}, cat states, and squeezed cat states~\cite{ref_Etesse, ref_Teh, ref_zheng, ref_yang}, have expanded the scope of quantum mechanical advantages across various technological applications.

\begin{figure}[b]
    \centering
    \includegraphics[width=0.9\linewidth]{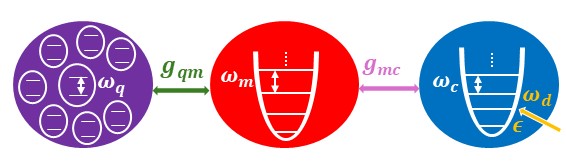}
    \caption{The schematic represents a tripartite hybrid quantum system. The purple disk denotes an ensemble of qubits, while the red disk signifies a mechanical resonator mode of frequency $\omega_m$. The blue disk corresponds to a single-mode cavity with resonance frequency $\omega_c$. An external drive, characterized by strength $\epsilon$ and frequency $\omega_d$, is represented by the yellow arrow within the blue disk. The central white circle inside the purple disk illustrates an individual qubit, modeled as a two-level quantum system with a transition energy $\hbar\omega_q$ between the ground state $\ket{g}$ and the excited state $\ket{e}$. Each qubit interacts with the mechanical resonator via a coupling strength $g_{qm}$ to the mechanical resonator, while the mechanical resonator is coupled to the cavity mode with an interaction strength  $g_{mc}$.}
    \label{fig_1_schematic}
\end{figure}

The recent development of experimental hybrid quantum systems for the transfer of quantum states across different frequency regimes~\cite{ref_Clerk2020,ref_Lauk2020,ref_Xiang2013}, known as quantum transducers, has opened new opportunities to investigate a wide range of phenomena. Although quantum transduction has been successfully demonstrated in the weak coupling regime of a hybrid system~\cite{ref_Mirhosseini2020}, recent advances have also introduced experimental setups in which solid-state spins are coupled to nanomechanical oscillators~\cite{ref_Wang2025,ref_Rosenfeld2021,ref_Shandilya2021,ref_Karg2020}, offering new possibilities with diverse coupling strengths. Many existing studies on hybrid quantum systems use mechanical modes primarily as intermediaries for efficient transfer of quantum states~\cite{ref_Mirhosseini2020,ref_neumann_narang,ref_Zhong2022,ref_Zhong2022b,ref_Molinares2022}. However, the role of externally addressable subsystems in the state of the mechanical mode of the intermediary device has remained largely unexplored. Critically, this involves the state of the intermediary mode itself, which is shaped by its interactions with surrounding subsystems. Previous experimental and theoretical research generating non-classical mechanical states mainly explored bipartite systems, where qubits are coupled to mechanical modes~\cite{ref_Lo2015,ref_Monroe1996,ref_Bild2023,ref_Marti2024} or mechanical oscillators are coupled to cavities~\cite{ref_qian,ref_nation,ref_Lorch,ref_hauer,ref_wise_new}. Most of these studies have concentrated on steady-state properties, with limited attention given to transient dynamical behavior.

We consider a tripartite hybrid quantum system consisting of a collection of qubits coupled to a mechanical mode, which in turn interacts with an externally driven cavity. By integrating three components into the hybrid platform, the system not only offers additional control parameters, but also unlocks rich multimode protocols. Operating in the strong coupling regime, we explore the effects of the nonlinear interactions between the mechanical and cavity mode, as well as between the qubit and mechanical mode. By applying a time-dependent drive that is switched off beyond a critical time, we observe well-defined oscillatory dynamics. Focusing on transient dynamics, we identify quantum states of motion that exhibit non-Gaussian characteristics. Utilizing the Wigner negative region ratio to quantify the volume of the negative region in the Wigner quasi-probability distribution and quantum Fisher information (QFI), we analyze the control parameters in the qubit and cavity components that enhance the non-Gaussianity of quantum states from their initial Gaussian nature. Our results demonstrate that starting with coherent states in the mechanical and cavity degrees of freedom, a time-dependent external drive scheme in the presence of fixed time-independent interaction and frequency schemes can lead to the emergence of highly non-Gaussian mechanical states. Additionally, the phase in the initial qubit state as well as coupling strength to the qubit subsystem significantly influence the non-Gaussianity of the mechanical state, highlighting potential applications in quantum technologies.

\textit{System and Hamiltonian:} The tripartite hybrid quantum setup is illustrated in Fig.~\ref{fig_1_schematic}. The total Hamiltonian $\hat{H}$ of the system in a reference frame rotating with the frequency $\omega_d$ of the external drive reads
\begin{eqnarray}
\label{eq_ham_total}
\hat{H} = \hat{H}_{\text{0}} + \hat{H}_{\text{int}} + \hat{H}_{\text{drive}},
\end{eqnarray}
where $\hat{H}_{\text{0}}$, $\hat{H}_{\text{int}}$, and $\hat{H}_{\text{drive}}$ represent the Hamiltonians corresponding to the non-interacting, the interacting and the external drive acting on the cavity part, respectively. The non-interacting Hamiltonian $\hat{H}_\text{0}$ is given by
\begin{eqnarray}
\label{eq_ham_non-int}
\hat{H}_{\text{0}}=
\frac{\hbar \omega_q}{2}\sum_{j=1}^{N_q} (\hat{\sigma}_{j}^z + \hat{I}_j)+\hbar \omega_m  \hat{b}^{\dagger} \hat{b} -\Delta  \hat{a}^{\dagger} \hat{a},
\end{eqnarray}
where $\hat{\sigma}_{j}^z = |e\rangle_j\langle e| - |g\rangle_j\langle g|$ and $\hat{I}_{j} = |e\rangle_j\langle e| + |g\rangle_j\langle g|$, and $\hat{b}^\dagger$ ($\hat{a}^\dagger$) and $\hat{b}$ ($\hat{a}$) denote the creation and annihilation operators for phonons (photons) in the mechanical (cavity) mode, respectively. The first term in Eq.~(\ref{eq_ham_non-int}) describe the qubit system that consists of $N_q$ qubits with a transition energy of $\hbar\omega_q$ between the ground state $\ket{g}_j$ and the excited state $\ket{e}_j$ of the $j$th qubit. The second term describes the mechanical resonator with its energy $\hbar\omega_m$, while the third term represents the cavity mode with detuning $\Delta$. Here, $\Delta$ is defined as the difference between the drive frequency $\omega_d$ and single-mode cavity frequency $\omega_c$, i.e., $\Delta=\hbar(\omega_d-\omega_c)$.

The interaction Hamiltonian $\hat{H}_{\text{int}}$ reads:
\begin{eqnarray}
\label{eq_int_ham}
\hat{H}_{\text{int}} = g_{qm} \sum_{j=1}^{N_q}(\hat{\sigma}^+_j \hat{b} + \hat{\sigma}^-_j\hat{b}^{\dagger})+g_{mc} \hat{a}^{\dagger}\hat{a}(\hat{b} +\hat{b}^{\dagger}),
\end{eqnarray}
where $g_{qm}$ represents the coupling strength between the qubits and the mechanical mode, and $g_{mc}$ denotes the interaction strength between the mechanical and cavity modes.

Finally, the Hamiltonian describing the time-dependent external drive is given by
\begin{eqnarray}
\label{eq_cdrive_ham}
\hat{H}_{\text{drive}}(t) = \epsilon(t) (\hat{a}^{\dagger} + \hat{a}).
\end{eqnarray}
The time-dependent external cavity drive $\epsilon(t)$ that appeared in Eq.~(\ref{eq_cdrive_ham}) has the form $\epsilon(t)=\epsilon_0\Theta(t-t_c)$. It implies that the drive is active with strength $\epsilon_0$ up to a critical time $t_c$ and beyond that it drops to zero. The time-dependent drive scheme, resembling a boxcar function, enables the potential discovery of a resourceful mechanical state within transient dynamics, exhibiting a periodic nature.

\begin{figure}[b]
    \centering
    \includegraphics[width=0.9\linewidth]{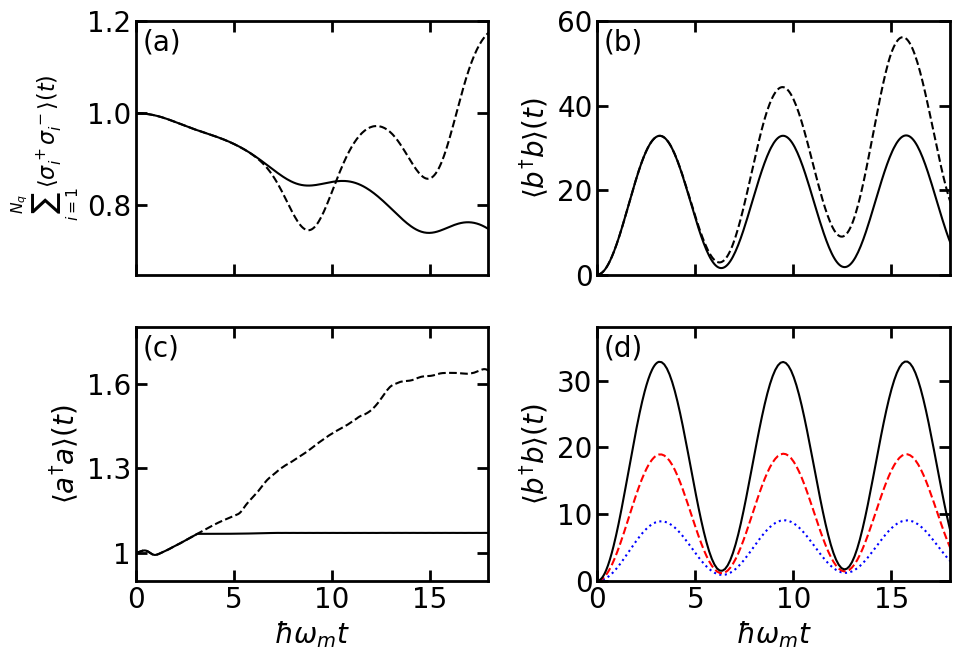}
    \caption{Populations of the three subsystems as a function of time $\hbar\omega_m t$ for $\ket{\psi(0)}=\frac{1}{\sqrt{2}}(\ket{eg}+\ket{ge})\otimes\ket{0}\otimes\ket{\alpha=1}$, $\omega_q/\omega_m =1$, $g_{qm}/\hbar\omega_m =0.05$, $\Delta/\hbar\omega_m =0$, and $\epsilon_0/\hbar\omega_m =0.3$. (a)--(c) The data shown correspond to qubit $\sum_{i=1}^{N_q}\braket{\hat{\sigma}^+_i\hat{\sigma}^-_i}(t)$, mechanical $\braket{\hat{b}^{\dagger}\hat{b}}(t)$ and cavity mode $\braket{\hat{a}^{\dagger}\hat{a}}(t)$ population, respectively, for $g_{mc}/\hbar\omega_m =2$, $\hbar\omega_m t_c = \pi$ (solid black line) and $\hbar\omega_m t_c \rightarrow \infty$ (dashed black line). (d) The population of the mechanical mode is plotted as a function of time for $g_{mc}/\hbar\omega_m =2$ (solid black line), $g_{mc}/\hbar\omega_m =1.5$ (dashed red line), and $g_{mc}/\hbar\omega_m =1$ (dotted blue line) using the critical time parameter $\hbar\omega_m t_c=\pi$.}
    \label{fig_2_pop_dynamics}
\end{figure}

\textit{Oscillatory dynamics}: We study the time evolution of the hybrid system governed by $\frac{d}{dt}\hat{\rho}_{\text{total}} = -\frac{i}{\hbar}[\hat{H},\hat{\rho}_{\text{total}}]$, starting from the initial density matrix $\hat{\rho}_{\text{total}}(0)=\ket{\psi(0)}\bra{\psi(0)}$. To specifically explore the properties of the mechanical mode, we trace out the qubit and cavity degrees of freedom from $\hat{\rho}_{\text{total}}$, resulting in the reduced density matrix $\hat{\rho}$ for the mechanical mode. We start by analyzing the time-dependent behavior of the qubit, mechanical, and cavity mode populations. The initial state considered in Fig.~\ref{fig_2_pop_dynamics} has the form $\ket{\psi(0)}=\frac{1}{\sqrt{2}}(\ket{eg}+\ket{ge})\otimes\ket{0}\otimes\ket{\alpha=1}$, i.e., the qubit part is in a symmetric superposition state, the mechanical mode is in its ground state, and the cavity mode is in a coherent state of amplitude $\alpha=1$. Figure~\ref{fig_2_pop_dynamics}(a)--(c) includes the population dynamics of the hybrid quantum system for two different drive schemes with parameters $\omega_q/\omega_m =1$, $g_{qm}/\hbar\omega_m =0.05$, $\Delta/\hbar\omega_m =0$, $g_{mc}/\hbar\omega_m =2$, and $\epsilon_0/\hbar\omega_m =0.3$. The dashed black line represents the scenario where the external cavity drive has a critical time $\hbar\omega_mt_c\rightarrow\infty$ whereas the solid black line corresponds to the case with $\hbar\omega_m t_c = \pi$. We observe that when the drive has a large critical time, both the photon population in the cavity and the phonon population in the mechanical mode continue to increase. Additionally, the qubit population gets enhanced as more qubit excitations are created. However, when the drive strength drops to zero at the critical time $\hbar \omega_m t_c = \pi$, the photon population ceases to increase and stabilizes at a constant value. In contrast, the qubit and phonon populations remain time-dependent. Notably, the phonon population exhibits a well-defined oscillatory behavior, where the oscillation amplitude has an upper threshold. This implies that beyond a certain maximal phonon occupation, higher modes do not become occupied throughout the evolution. Focusing on this oscillatory behavior, we note that the first peak appears at $\hbar \omega_m t = \pi$. Figure~\ref{fig_2_pop_dynamics}(d) illustrates the phonon mode population as a function of time using parameters $\omega_q/\omega_m =1$, $g_{qm}/\hbar\omega_m =0.05$, $\Delta/\hbar\omega_m =0$, $\epsilon_0/\hbar\omega_m =0.3$, and $\hbar\omega_m t_c =\pi$ for three different values of $g_{mc}/\hbar\omega_m$. The dotted blue line, dashed red line, and solid black line correspond to the values $g_{mc}/\hbar\omega_m =1, 2,$ and $3$, respectively. We see that the oscillation frequency of the phonon mode population remains unchanged. However, as $g_{mc}/\hbar\omega_m$ increases, the amplitude of oscillation also increases. It is because when the drive is activated during the period $t\leq t_c$, a larger value of $g_{mc}/\hbar\omega_m$ allows for the generation of more excitations in the phonon mode through phonon-photon coupling, leading to an increase in the maximal phonon mode population over time. It is worth noting that while the mechanical mode starts in the ground state at $t=0$, it acquires a finite non-zero population of $\braket{\hat{b}^{\dagger}\hat{b}}\approx2$ at later times, even when oscillatory dynamics bring the population to a minimum.

\begin{figure}[t]
    \centering
    \includegraphics[width=0.95\linewidth]{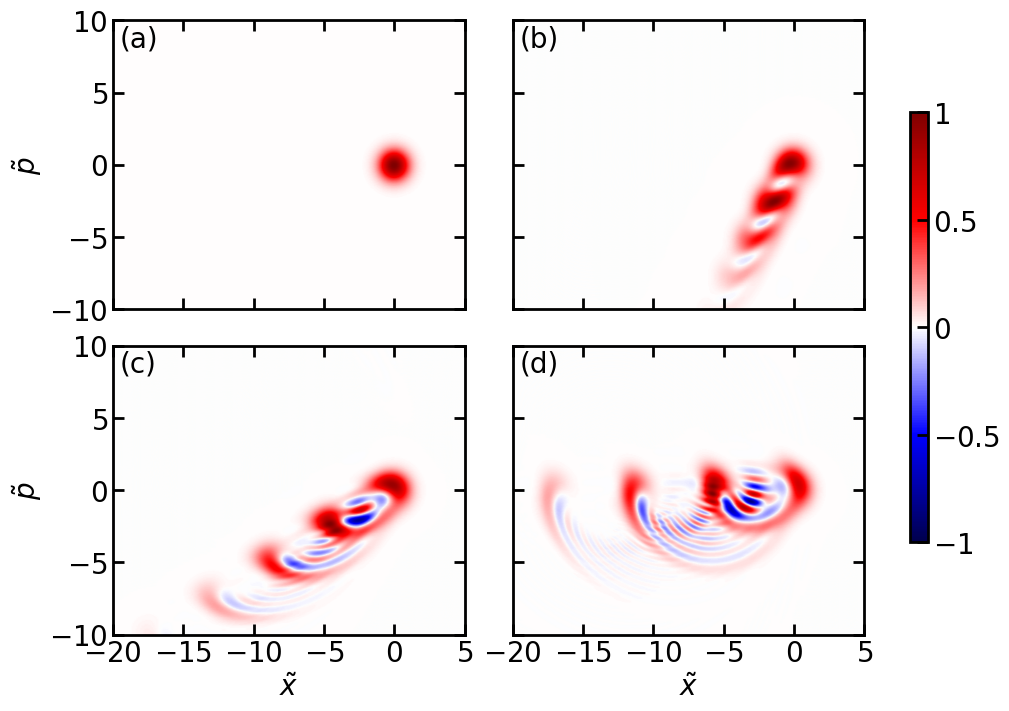}
    \caption{Snapshots of $W(x,p)/W_{max}$ for $\omega_q$/$\omega_m = 1$, $g_{qm}$/$\hbar\omega_m= 0.05$, $g_{mc}$/$\hbar\omega_m$=$2$, $\epsilon_0$/$\hbar\omega_m$=$0.5$, and $\hbar\omega_m t_c=\pi$ as functions of dimensionless $x$ and $p$ co-ordinate for the initial state $\frac{1}{\sqrt{2}}(\ket{eg}+{ge})\otimes\ket{0}\otimes\ket{\alpha=1}$. (a)--(d) correspond to $\hbar\omega_m t=0$, $\pi/3$, $2\pi/3$, and $\pi$, respectively.}
    \label{fig_3_wigdist_oscillations}
\end{figure}

\begin{figure}[ht]
    \centering
    \includegraphics[width=0.7\linewidth]{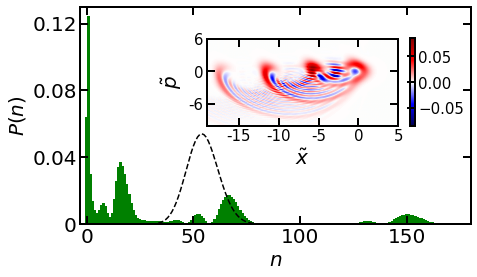}
    \caption{Fock space distribution $P(n)$ of the mechanical mode state calculated at $\hbar\omega_m t=\pi$ using the parameters $\omega_q$/$\omega_m = 1$, $g_{qm}$/$\hbar\omega_m= 0.05$, $g_{mc}$/$\hbar\omega_m$=$2$, $\epsilon_0$/$\hbar\omega_m$=$0.8$, and $\hbar\omega_m t_c=\pi$ as functions of fock space index $n$ for the initial state $\frac{1}{\sqrt{2}}(\ket{eg}-\ket{ge})\otimes\ket{0}\otimes\ket{\alpha=1}$. The inset shows the Wigner quasi-probability distribution in the scaled phase space for the corresponding state.}
    \label{fig_4_wig-fock_ng}
\end{figure}

To further explore these dynamics, we analyze the time-dependent Wigner quasi-probability distribution. Figure 3 illustrate the scaled Wigner quasi-probability distribution $W(x,p)/W_{max}$ as a function of the dimensionless phase-space coordinates at four different time instances. Since $W_{max}$ varies for each subplot, we use the scaled Wigner distribution to simplify visualization. The Wigner quasi-probability distribution for the mechanical mode, which evolves from the ground state, exhibits negative regions in its phase space (highlighted by the blue shaded area in Fig.~\ref{fig_3_wigdist_oscillations}(c)--(d)) indicating the emergence of non-Gaussian characteristics. The negative regions do not appear immediately; as seen in Fig.~\ref{fig_3_wigdist_oscillations}(b), the distribution initially lacks any blue shaded areas, indicating a finite time is required for their formation. Figure~\ref{fig_3_wigdist_oscillations}(d) and the inset of Fig.~\ref{fig_4_wig-fock_ng} prominently display these negative areas, emphasizing the non-Gaussian nature of the state observed at the critical time. Furthermore, Fig.~\ref{fig_4_wig-fock_ng} presents the Fock space distribution of the mechanical state calculated at the critical time using the parameters: $\omega_q/\omega_m = 1$, $g_{qm}/\hbar\omega_m= 0.05$, $g_{mc}/\hbar\omega_m= 2$, $\epsilon_0/\hbar\omega_m= 0.8$, and $\hbar\omega_m t_c=\pi$ for the initial state $\frac{1}{\sqrt{2}}(\ket{eg}-\ket{ge})\otimes\ket{0}\otimes\ket{\alpha=1}$. This distribution exhibits a structure that includes contributions from both low-energy and high-energy Fock states. Specifically, the components $n=1$ and $n=0$ contribute $12.4\%$ and $6.4\%$, respectively, along with contributions from higher values of $n$. Additionally, the black dashed line in Fig.~\ref{fig_4_wig-fock_ng} represents a coherent Gaussian state with a similar amplitude, i.e., $\braket{\hat{b}^{\dagger}\hat{b}}\approx 54$, which serves to highlight the differences between a typical coherent Gaussian state and the non-Gaussian state emerging in this hybrid quantum system.

\textit{Negative volume in Wigner distribution}: To quantify the non-Gaussian nature of the mechanical state under study, we use the Wigner negative region ratio~\cite{ref_Marti2024,ref_wise_new}, defined as 
\begin{equation}
    \zeta = \frac{\int |W_-(x,p)| dx dp}{\int W_+(x,p)dx dp},
\end{equation}
where $W_-$ and $W_+$ correspond to the negative and positive values of Wigner distribution respectively. For a Gaussian state, in absence of negative regions in Wigner distribution, the value of $\zeta$ is zero. As a reference for non-Gaussian state, a cat state of the form $\frac{1}{\sqrt{2(1-\exp{(-2|\beta|^2)})}}(\ket{\beta}-\ket{-\beta})$ has a $\zeta$ value of $0.23$ for an amplitude $\beta=1$. The state shown in Fig.~\ref{fig_4_wig-fock_ng} has larger $\zeta$ value of $0.32$. Moreover, in our system, the value of $\zeta$ is influenced by various control parameters, demonstrating the tunability of the hybrid system.

\begin{figure}[b]
    \centering
    \includegraphics[width=0.95\linewidth]{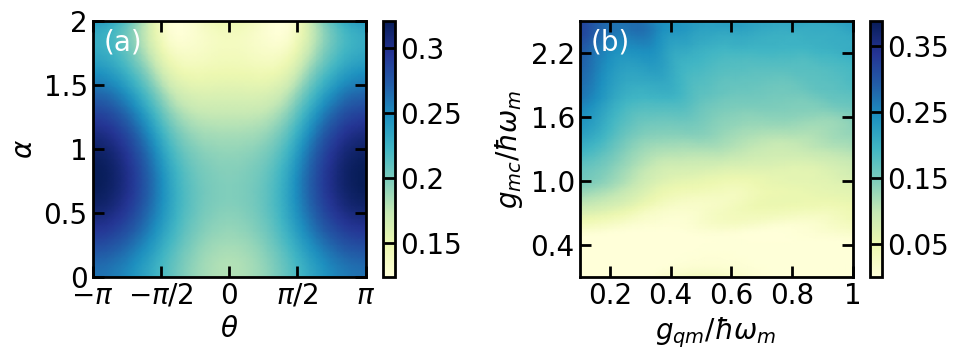}
    \caption{Contour plots show the dependence of $\zeta$ on qubit and cavity control parameters $\theta$, $\alpha$, $g_{qm}/\hbar\omega_m$, and $g_{mc}/\hbar\omega_m$ for $\omega_q$/$\omega_m$ = 1, $\Delta/\hbar\omega_m=0$, $\epsilon_0$/$\hbar\omega_m$ = 0.8, and $\hbar\omega_m t_c =\pi$. (a) $\zeta$ is shown as a function of $\theta$ and $\alpha$ for the initial state $\ket{\psi(0)}=\frac{1}{\sqrt{2}}(\ket{eg}+\exp(\iota\theta)\ket{ge})\otimes\ket{0}\otimes\ket{\alpha}$ with $g_{qm}$/$\hbar\omega_m$ = 0.05 and $g_{mc}/\hbar\omega_m=2$. (b) The dependence of $\zeta$ on $g_{qm}$/$\hbar\omega_m$ and $g_{mc}/\hbar\omega_m$ for the initial state $\ket{\psi(0)}=\frac{1}{\sqrt{2}}(\ket{e}+\ket{g})\otimes\ket{0}\otimes\ket{\alpha=1}$.}
    \label{fig_5_ncr_tuning}
\end{figure}

Figure~\ref{fig_5_ncr_tuning}(a) shows $\zeta$ as a function of phase $\theta$ encoded in the qubit component of the initial state and the amplitude $\alpha$ of the coherent state in cavity mode. It can be seen that for a fixed $\theta$, as $\alpha$ increases from $0$ to approximately $0.8$, $\zeta$ also increases, reaching a peak before subsequently decreasing as $\alpha$ continues to grow. When $\alpha$ exceeds $1$, initial quantum correlations in the cavity mode starts diminishing, leading to a mechanical state at the critical time with reduced non-Gaussian characteristics. The initial increase in $\zeta$ when $\alpha$ rises from $0$, can be attributed to the presence of a greater number of Fock modes in the cavity component of the initial state. This, in turn, enables the cavity drive to couple multiple cavity states, resulting in stronger correlation between the mechanical mode and the cavity modes at later times. Conversely, for a fixed $\alpha$, $\zeta$ reaches its maximum value at $\theta=\pm \pi$ highlighting the qubit-controlled nature of the system and its corresponding states. In particular, the phase $\theta$ directly affects the oscillation amplitude in the population dynamics of the mechanical mode, with $\theta=\pm \pi$ leading to the maximum amplitude while keeping all other parameters unchanged. 

In Fig.~\ref{fig_5_ncr_tuning}(b), $\zeta$ is shown as a function of coupling strengths $g_{qm}$/$\hbar\omega_m$ and $g_{mc}/\hbar\omega_m$ for the initial state $\ket{\psi(0)}=\frac{1}{\sqrt{2}}(\ket{eg}+\ket{ge})\otimes\ket{0}\otimes\ket{\alpha=1}$ with parameters $\omega_q$/$\omega_m$ = 1, $\Delta/\hbar\omega_m=0$, $\epsilon_0$/$\hbar\omega_m$ = 0.8, and $\hbar\omega_m t_c =\pi$. We see that a larger value of $g_{mc}/\hbar\omega_m$ results in an increased $\zeta$. Specifically, when $g_{qm}$/$\hbar\omega_m$ is fixed, $\zeta$ remains negligible for $g_{mc}$/$\hbar\omega_m$ values roughly $0.6$. This range of coupling strength corresponds to a scenario in which the second interaction term in Eq.~(\ref{eq_int_ham}) can be linearized making the non-linear effects of the interaction between the mechanical and cavity modes insignificant. This observation highlights the crucial role of nonlinearity in generating non-Gaussian characteristics in mechanical mode.

\begin{figure}
    \centering
    \includegraphics[width=0.95\linewidth]{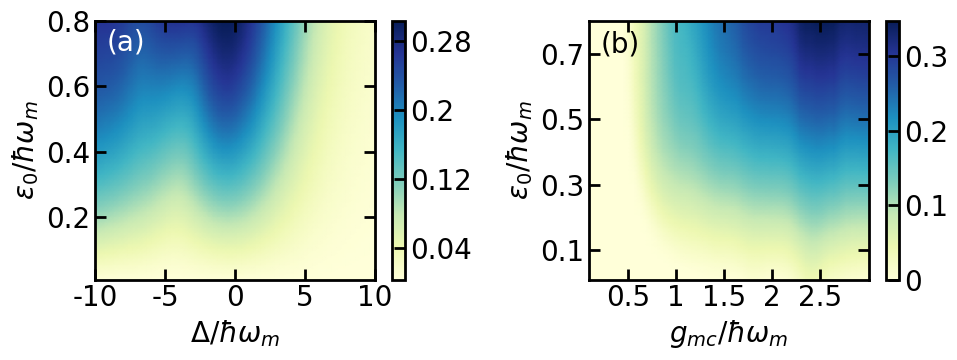}
    \caption{Contour plots of $\zeta$ as a function of cavity control parameters $g_{mc}/\hbar\omega_m$, $\Delta/\hbar\omega_m$, and $\epsilon_0/\hbar\omega_m$ for $g_{qm}/\hbar\omega_m$ = 0.05, and $\hbar\omega_m t_c =\pi$. (a) $\zeta$ is plotted as a function of $\Delta/\hbar\omega_m$ and $\epsilon_0/\hbar\omega_m$ for the initial state $\ket{\psi(0)}=\frac{1}{\sqrt{2}}(\ket{e}+\ket{g})\otimes\ket{0}\otimes\ket{\alpha=1}$ with $g_{mc}/\hbar\omega_m=2$. (b) The dependence of $\zeta$ on $\epsilon_0$/$\hbar\omega_m$ and $g_{mc}/\hbar\omega_m$ for the initial state $\ket{\psi(0)}=\frac{1}{\sqrt{2}}(\ket{e}+\ket{g})\otimes\ket{0}\otimes\ket{\alpha=1}$ with $\Delta=0$.}
    \label{fig_5_ncr_tuning_new}
\end{figure}

Figure~\ref{fig_5_ncr_tuning_new}(a) shows the dependence of $\zeta$ on the detuning parameter $\Delta/\hbar\omega_m$ and the drive strength $\epsilon_0/\hbar\omega_m$. Given the parameters $\omega_q/\omega_m=1$, $g_{qm}/\hbar\omega_m = 0.05$, and $\hbar\omega_m t_c =\pi$, the results suggest that a non-positive detuning combined with a stronger external drive enhances the value of $\zeta$. When $\epsilon_0/\hbar\omega_m$ is below $0.2$, $\zeta$ remains negligibly small for all detuning values. However, for higher values of $\epsilon_0/\hbar\omega_m$, $\zeta$ reaches its peak at $\Delta=0$. Figure~\ref{fig_5_ncr_tuning_new}(b) illustrates $\zeta$ as a function of the drive strength $\epsilon_0/\hbar\omega_m$ and the mechanical-cavity coupling strength $g_{mc}/\hbar\omega_m$ at zero detuning, for the initial state $\ket{\psi(0)}=\frac{1}{\sqrt{2}}(\ket{e}+\ket{g})\otimes\ket{0}\otimes\ket{\alpha=1}$. It is evident that increasing $g_{mc}/\hbar\omega_m$ and $\epsilon_0/\hbar\omega_m$ leads to a larger value of $\zeta$.

\begin{figure}[ht]
    \centering
    \includegraphics[width=0.95\linewidth]{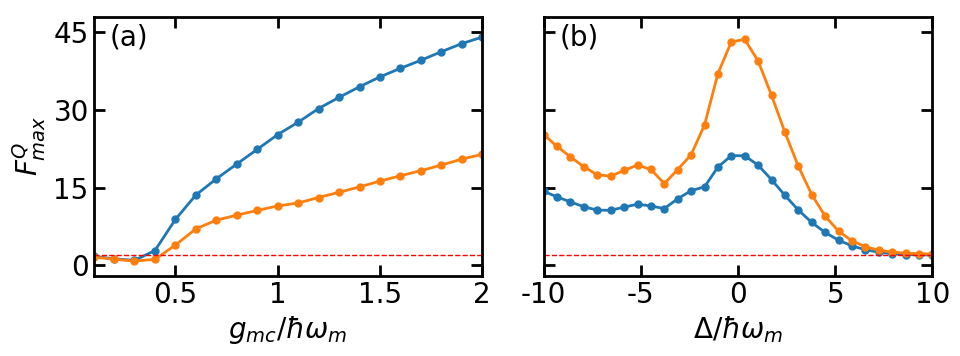}
    \caption{$F_{max}^{Q}$ of the mechanical mode is plotted as a function of (a) mechanical-cavity coupling $g_{mc}/\hbar\omega_m$ at fixed detuning $\Delta/\hbar\omega_m=0$, and (b) detuning $\Delta/\hbar\omega_m$ at fixed coupling $g_{mc}/\hbar\omega_m=2$. The system is initialized in the state $\ket{\psi(0)}=\frac{1}{\sqrt{2}}(\ket{eg}+\exp{(\iota\theta)}\ket{ge})\otimes\ket{0}\otimes\ket{\alpha=1}$, with parameters $\omega_q/\omega_m =1$, $g_{qm}/\hbar\omega_m =0.05$, $\epsilon_0/\hbar\omega_m =0.8$, and $\hbar\omega_m t_c=\pi$. The solid blue and solid orange lines correspond to $\theta=0$ and $\pi$, respectively. The dashed red line represents a coherent mechanical state with amplitude $\beta$.}
    \label{fig_6_qfi}
\end{figure}

\textit{Quantum Fisher information}: In order to further characterize the state in the intermediary mechanical mode of the hybrid quantum system, we calculate the quantum Fisher information. The quantum Fisher information associated with a density matrix $\hat{\rho}$ and an observable $\hat{G}$ is given by~\cite{ref_Marti2024,ref_qfi_lloyd}
\begin{eqnarray}
    F^{Q}[\hat{\rho},\hat{G}] = 2\sum_{k,l,\eta_l+\eta_k>0}\frac{(\eta_k-\eta_l)^2}{\eta_l+\eta_k}|\braket{k|\hat{G}|l}|^2,
    \label{eq_QFI}
\end{eqnarray}
where $\ket{k}$ and $\eta_k$ are the eigenstates and corresponding eigenvalues of the density matrix $\hat{\rho}$. We define a generalized displacement amplitude operator as $\hat{G}(\phi)=\hat{X}\sin(\phi)+\hat{P}\cos(\phi)$ where $\phi$ represents the angular direction. To perform parameter estimation for a given state $\hat{\rho}$, we calculate the maximum value of $F^{Q}$ given by $F_{max}^{Q}[\hat{\rho}]=\text{max}\{ F^{Q}[\hat{\rho},\hat{G}(\phi)]; \phi\in [0,2\pi]\}$.

Figure~\ref{fig_6_qfi} shows the behavior of $F_{\text{max}}^{Q}$ as a function of detuning $\Delta/\hbar\omega_m$ (Fig.~\ref{fig_6_qfi}(a)) and cavity-mechanical coupling strength $g_{mc}/\hbar\omega_m$ (Fig.~\ref{fig_6_qfi}(b)), for two different values of the phase $\theta$ encoded in the qubit part of the initial state $\ket{\psi(0)}=\frac{1}{\sqrt{2}}(\ket{eg}+\exp{(\iota\theta)}\ket{ge})\otimes\ket{0}\otimes\ket{\alpha=1}$. The parameters $\omega_q/\omega_m =1$, $g_{qm}/\hbar\omega_m =0.05$, $\epsilon_0/\hbar\omega_m =0.8$, and $\hbar\omega_m t_c=\pi$ are kept constant across both subplots. It can be seen that phase $\theta = \pi$ (solid blue line) results in significantly larger values of $F_{\text{max}}^{Q}$ compared to $\theta = 0$ (solid yellow line). This enhancement results from the increase in the population of the mechanical mode. In other words, when $\theta = \pi$, the higher-energy Fock states of the mechanical mode become more populated compared to the case when $\theta = 0$. This underscores the importance of the qubit state phase $\theta$ as a control parameter. 

In Fig.~\ref{fig_6_qfi}(a), we see that for $g_{mc}/\hbar\omega_m < 0.5$, the value of $F_{\text{max}}^{Q}$ remains below or at most equal to that of a coherent state with any amplitude $\beta$ (indicated by the dashed red line). This region corresponds to the weak-coupling regime, where the interaction Hamiltonian between the mechanical and cavity modes can be linearized to accurately describe the system characteristics. However, as $g_{mc}/\hbar\omega_m$ increases, $F_{\text{max}}^{Q}$ also increases, highlighting the importance of the strong coupling regime. In this regime, nonlinearity becomes prominent, enabling the formation of non-Gaussian quantum states with enhanced QFI. Figure~\ref{fig_6_qfi}(b) shows that $F_{\text{max}}^{Q}$ reaches its peak at zero detuning. As $\Delta/\hbar\omega_m$ increases on the positive side of the axis, $F_{\text{max}}^{Q}$ gradually decreases, eventually converging to the characteristic value of a coherent state at a large positive detuning. On the other hand, for negative detuning, $F_{\text{max}}^{Q}$ initially decreases but then increases again as $\Delta/\hbar\omega_m$ becomes more negative. In the regime of large positive detuning, the cavity mode effectively decouples from the mechanical mode, and the system evolves as though the external drive is absent. Conversely, for negative detuning, the drive actively mediates coupling between the two modes, facilitating the generation of non-Gaussian states with large QFI (see Fig.~\ref{fig_5_ncr_tuning_new}(a) for Wigner negative region ratio characterization). This behavior emphasizes the pivotal role of detuning as a key control parameter in the system.

\textit{Conclusion}: We demonstrate that a time-dependent driving protocol resembling a boxcar function gives rise to well-defined oscillatory behavior, wherein the average population of the mechanical mode remains confined within a specific amplitude threshold. For a particular range of parameters, the system displays a periodic recurrence of the mechanical mode state that initially appears at $\hbar\omega_m t=\pi$. This recurring state is marked by a high Wigner negative region ratio and large quantum Fisher information, both signifying its non-Gaussian quantum character. Notably, the system begins with the mechanical mode in its ground state and the cavity mode in a coherent state, both of which are Gaussian semi-classical in nature. Our results emphasize the pivotal role of tunable qubit parameters, specifically the phase and qubit-mechanical interaction strength, in facilitating and controlling the development of non-Gaussian features in the mechanical mode. These observations assume that losses in various parts of the hybrid quantum system are negligible and are thus idealized, but are generally supported by the operational timescales demonstrated in several cutting-edge experiments.

\textit{Acknowledgment:} We gratefully acknowledge funding from the National Science Foundation (NSF) under NSF Award 2137828 ``QuIC-TAQS: Deterministically Placed Nuclear Spin Quantum Memories for Entanglement Distribution", NSF Award 2246394 ``CAREER: First Principles Design of Error-Corrected Solid-State Quantum Repeaters", and NSF Award 2107265 ``U.S.-Ireland R\&D Partnership: Collaborative Research: CNS Core: Medium: A unified framework for the emulation of classical and quantum physical layer networks".

\end{document}